\documentclass[twocolumn,prl,aps,amsfonts]{revtex4}
\usepackage{graphicx}
\usepackage{psfrag, color}
\usepackage{verbatim}

\begin{document}

\title{Impact of spin-orbit coupling on quantum Hall nematic phases}

\author{M. J. Manfra$^1$,  R. de Picciotto$^1$, Z. Jiang$^2$, S. H. Simon$^1$, L. N. Pfeiffer$^1$, K. W. West$^1$, and A. M. Sergent$^1$}

\address{
 ${}^1$ Bell Laboratories, Alcatel-Lucent, Murray Hill, New Jersey
 \\ ${}^2$ NHMFL, Florida State University, Tallahassee, Florida}

\begin{abstract}
Anisotropic charge transport is observed in a two-dimensional
(2D) hole system in a perpendicular magnetic field at filling
factors $\nu$=7/2, $\nu$=11/2, and $\nu$=13/2 at low temperature.  In stark contrast, the transport at $\nu$=9/2 is {\it isotropic} for
all temperatures.  Isotropic hole transport at $\nu$=7/2 is restored for sufficiently low 2D densities or an asymmetric confining potential.  The density and symmetry dependences of the observed anisotropies suggest that strong spin-orbit coupling in the hole system contributes to the unusual transport behavior.
\end{abstract}

\maketitle

Presently, the study of the properties of half-filled
Landau levels of clean two-dimensional (2D) systems is the
focus of intense research.  In 2D {\it electron} systems
(2DESs) at half-filling, a surprisingly diverse set of ground
states has been uncovered.   In the N=0 Landau level (LL), at $\nu=1/2$
and $\nu=3/2$, compressible composite-fermion Fermi liquid states
are observed \cite{willett1,Olle}.  Here $\nu=hcn/eB$ is the filling
factor with $B$ the magnetic field, and $n$ the carrier density.
In the N=1 Landau level, DC transport measurements have
convincingly demonstrated the presence of incompressible quantized
Hall states at $\nu$=5/2 and $\nu$=7/2 \cite{willett3,Pan1,Jim1}.  At half-filling in the $ 2 \leq $ N $\stackrel{<}{\sim} 5$ Landau
levels, electronic transport is anisotropic \cite{mike1,Ru1} and is
consistent with either a quantum smectic or nematic phase (i.e., ``striped" phase)\cite{Kivelson}.  While it is clear that all of these
phenomena derive from strong electron-electron interactions, the
exact relationship between the different ground states possible in
half-filled LLs and the sample parameters necessary to
stabilize one phase over another remain interesting experimental questions \cite{Pan2,mike2,rezayi1}.  Access to a greater range of sample parameters than is currently available in high mobility 2DESs may enhance our understanding of these exotic states.

Transport studies of high mobility two dimensional {\it hole}
systems (2DHSs) in GaAs offer a complimentary approach to the investigation
of correlation physics in 2D systems \cite{shayegan}.  The larger effective mass
of holes (m$_h$$\sim$0.5 vs. m$_e$$\sim$0.067, in units of the
free electron mass) reduces kinetic energy such that interactions play a more promiment role at a
given 2D density.  In addition, 2DHSs offer a ideal platform to study the impact of spin-orbit coupling on half-filled Landau levels
since spin-orbit coupling can significantly alter the ground state through mixing
of the light and heavy hole states \cite{roland1}.

In this Letter we detail the impact of strong spin-orbit coupling on quantum Hall nematic phases.  We present low temperature magnetotransport
measurements of a series of extremely high mobility, carbon-doped, 2DHSs
grown on the high symmetry (100) surface of GaAs.  At T $\sim$ 15mK we observe a pronounced
anisotropy in transport at filling factors $\nu$=7/2, $\nu$=11/2, and $\nu$=13/2 while the transport at $\nu$=9/2 remains {\it
isotropic}.  The resistance at $\nu$=7/2 in the [01$\bar{1}$]
direction exceeds the resistance in the [011] direction by
a factor of $\sim $10$^4$ \cite{Simon}. The observed transport anisotropies are extremely sensitive
to temperature. Isotropic transport is restored for
temperatures greater than T$\sim$130mK. Furthermore, isotropic transport at $\nu$=7/2 can also be restored by reducing the density of the 2DHS in a quantum well or by changing the symmetry of the potential confining the 2DHS at a constant 2D density. Our results for a 2D hole system
differ substantially from 2D electron transport where an isotropic fractional quantum Hall state
is observed at $\nu$=7/2, and the strongest anisotropy occurs at
$\nu$=9/2.

\begin{table}
\begin{center}
\begin{tabular}{|c|ccc|}
\hline
sample & p (10$^{11}$cm$^{-2}$) & $\mu$ (10$^6$cm$^2$/Vs) & structure  \\
\hline
  A & 1.2 & 2.0 & 20nm QW  \\
  B & 2.0 & 1.5 & 20nm QW  \\
  C & 2.3 & 1.3 & 20nm QW  \\
  D & 2.3 & 0.8 & SHJ  \\
\hline
\end{tabular}
\end{center}
\caption{Sample parameters of the 4 carbon-doped (100) 2DHSs studied in this work.  p is the 2D density and $\mu$ is the mobility.  QW indicates a quantum well, while SHJ indicates a single heterojunction.}
\end{table}

\begin{figure}
\includegraphics[width=\columnwidth]{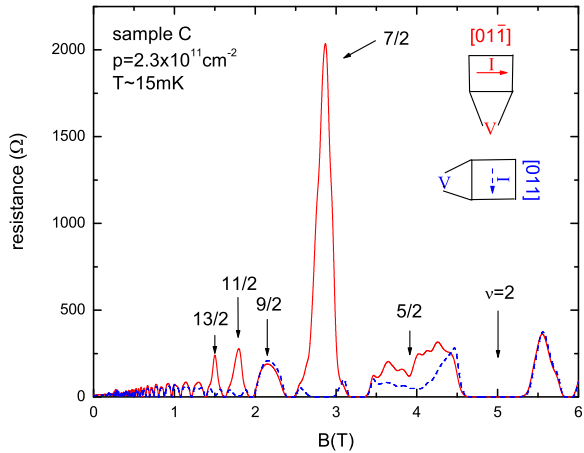}
 \caption{(Color online) Overview
of magnetoresistance in sample C with p=2.3$\times$10$^{11}$cm$^{-2}$ at T $\sim$ 15mK.  The dashed trace is measured with current flowing
in the [011] direction.  The solid line trace corresponds to
the magnetoresistance with the current flowing in the [01$\bar{1}$]
direction.  The transport is anisotropic at $\nu$=7/2, $\nu$=11/2, and $\nu$=13/2 but remains isotropic at $\nu$=9/2.}
\end{figure}

Square samples (4mm by 4mm) from a total of 4 separate wafers were examined in this study.  All samples are grown on the (100) surface of GaAs by molecular beam epitaxy \cite{manfra}.  The sample parameters are detailed in Table I.  Samples A, B, and C are 20nm wide, symmetrically doped, GaAs/AlGaAs
quantum wells.  The samples have a fixed setback of 80nm and the Al mole fraction of the barrier is varied to control the density.  Sample D is a single heterojunction.  The mobility of sample A reaches 2$\times$10$^6$cm$^2$/Vs at low temperature, which combined with a large effective mass in these samples (m$_{h}$$\sim$0.54) \cite{Han}, attest to the unprecedented quality of these newly developed structures.  Equally important, the zero field
mobility anisotropy  for these (100) structures is $\leq$ 20\% \cite{manfra}.  This residual anisotropy is also typical
in high mobility 2DESs.  Transport is measured in two separate dilution refrigerators which reach base temperatures of T$\sim$45mK and T$\sim$15mK. Standard low frequency lock-in techniques with excitation currents $\leq$10nA are used to monitor the resistance.

\begin{figure}
\includegraphics[width=.9\columnwidth]{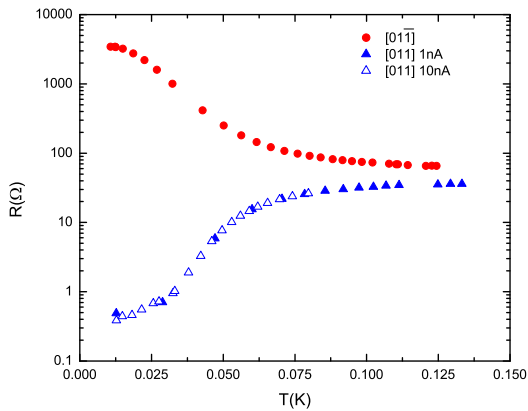}
 \caption{(Color online) Magnetoresistance along [01$\bar{1}$] (solid circles) and [011] (open and solid triangles)
directions as a function of temperature for sample C at $\nu$=7/2.  Along the [01$\bar{1}$] direction, a 1nA excitation is used.  Along the [011] direction, the resistance is measured with 1nA (solid triangles) and 10nA (open triangles) excitations.}
\end{figure}

Fig. 1 presents an overview of transport in sample C, for filling
factors $\nu \geq$ 2 at T$\sim$15mK along the [011] and
[01$\overline{1}$] directions.  We note that the resistances along [011] and [01$\overline{1}$] have {\it not}
been scaled to have equal amplitude at low magnetic field.  Starting from the low field regime, the resistance becomes clearly anisotropic at $\nu$=13/2, and $\nu$=11/2, with weaker features visible at $\nu$=15/2 and $\nu$=17/2. In sharp contrast, the
resistance is {\it isotropic} at $\nu$=9/2.  At
$\nu$=7/2 the resistance again becomes highly anisotropic.  No indication of a quantized Hall state in R$_{xy}$ is present at $\nu$=7/2 (not shown).  Moreover, the resistance ratio, R$_{[01\bar{1}]}$/R$_{[011]}$,
reaches $\sim$ 10$^{4}$ at $\nu$=7/2.  This behavior is strongly reminescent of the anisotropic transport first seen in 2D electron systems at half filling, but only at $\nu$ $\geq$ 9/2 in the N$\geq$2 LLs.  The appearance of such highly anisotropic transport has been interpreted as evidence for the formation of a unidirectional charge density wave (i.e., ``striped") phase or nematic liquid crystal-like phase in 2D electron systems \cite{Kivelson}. The data of Fig. 1 suggests that similar physics may be active in our hole system at $\nu$=7/2, $\nu$=11/2, and $\nu$=13/2.  While we have not yet systematically studied $\nu$=5/2, it shows no strong anisotropy in this sample.  The temperature evolution of the magnetotransport for filling
factor $\nu$=7/2 for sample C is shown in Figure 2.  At
T=130mK, the resistance is nearly
isotropic.  Upon reducing the temperature below
T=80mK, the anisotropy at $\nu$=7/2 develops rapidly in a manner similar to that seen 2D electron systems at $\nu$=9/2 \cite{Ru1,mike1}.

To date, only one other study has been dedicated to the investigation of
half-filled Landau levels in 2DHSs. Shayegan {\it et al.} \cite{mansour1} have reported intriguing anisotropic transport at half-filling for 2DHSs grown on the (311)A orientation, but the exploration of anisotropic behavior in excited hole Landau levels of (311)A samples has been
hindered by the presence of a significant transport anisotropy at
zero magnetic field \cite{Heremans}.
The data of Fig. 1 unambiguously demonstrates that 2DHSs grown on the (100) orientation of GaAs can display a novel sequence of isotropic/anisotropic states at half-filling while maintaining isotropic behavior at zero magnetic field.

The observation of isotropic transport for 2D holes at $\nu$=9/2
flanked by strongly anisotropic transport at $\nu$=11/2 and $\nu$=7/2 in sample C is
the most striking feature of this study.  In 2D electron systems
which display anisotropic transport, the anisotropy only resides
in the N$\geq$2 LL's and also shows the largest resistance ratio
at $\nu$=9/2.  Although 2D electron systems do not exhibit anisotropic transport in the
N=1 Landau level at $\nu$=5/2 and $\nu$=7/2 in a perpendicular
magnetic field, Lilly \cite{mike2} and Pan
\cite{Pan2} have observed that the incompressible quantum Hall
states at $\nu$=7/2 and $\nu$=5/2 are replaced by compressible
anisotropic states under the application of large in-plane magnetic fields.
These results suggest that the physics influencing
the formation of compressible striped phases in the N$\geq$2 LL's
may be active in the N=1 LL under the appropriate change of the
effective interaction induced by the large in-plane field. In
numerical studies, Rezayi and Haldane (RH) \cite{rezayi1} have
shown that the incompressible quantum Hall state at $\nu$=5/2 is
near a phase transition into a compressible striped phase. Similar behavior may be expected at $\nu$=7/2.  In
the pseudopotential formulation of the FQHE \cite{Haldane1}, the
nature of the ground state is found to depend sensitively on the
relative strengths of the pseudopototential parameters $V_1$ and
$V_3$, where $V_m$ is the energy of a pair of electrons in a state
of relative angular momentum $m$.  RH find that at $\nu$=5/2,
small variations in $V_1$ and $V_3$ can drive the phase transition
and suggest that the proximity of the critical point to the
Coulomb potential is the principle reason that transport becomes
anisotropic in the tilting experiments.

What distinguishes our 2D hole system from the 2D electron system
such that the fractional quantum Hall state at $\nu$=7/2 is
destabilized and replaced by an compressible anisotropic state and
the transport at $\nu$=9/2 remains isotropic rather than
displaying the anisotropy seen in electron systems?  We suggest that
the strong spin-orbit coupling in the 2DHS is a critical
difference.   Spin-orbit coupling strongly mixes valence band
states, which alters the orbital structure of hole Landau levels at
B$\neq$0 \cite{Yang1, Ekenberg1}.  The nature of the single
particle wavefunctions that comprise a given LL alters the
pseudopotential parameters, significantly influencing the
correlations among the holes \cite{MacDonald1,Yang2}.  Following
Ref. \cite{Yang2}, we have self-consistently calculated the Landau
level structure in the Hartree approximation (while keeping axial
terms as in \cite{Yang1}).  The energy level and orbital structure of the valence LLs appropiate to our samples are, not suprisingly, quite complex, but agree at least qualitatively with many of our observations. For sample C, at $\nu$=7/2, the valence LL is a mixture of primarily the N=2, N=4, and N=5 orbitals, which results in a hole-hole interaction potential consistent with a striped phase.  At $\nu$=9/2, the valence LL is a mixture of N=1, N=2 and N=3.  The N=1 LL contributes 30\% to the wavefunction while the N=2 and N=3 levels contribute a total of 50\%.  As the data of Fig. 1 show, the inclusion of the N=1 LL appears to drive $\nu$=9/2 into an isotropic state.
\begin{figure}
\includegraphics[width=\columnwidth]{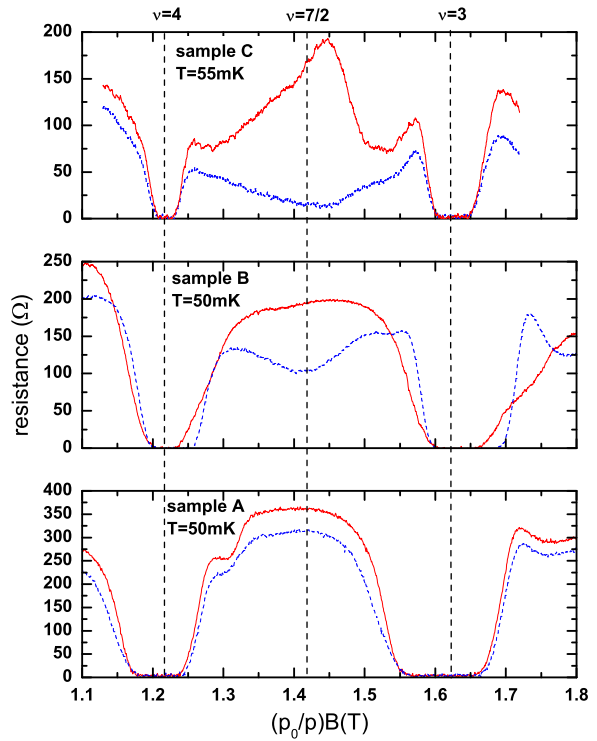}
 \caption{(Color online) Density dependence of the resistance anisotropy in the vicinity of $\nu$=7/2 at T $\sim$ 50mK for samples A, B, and C.  The magnetic field axis for each sample has been scaled by the density of sample A, $p_0$=1.2$\times$10$^{11}$cm$^{-2}$, to facilatate comparison at $\nu$=7/2. As the density is reduced from 2.3x10$^{11}$cm$^{-2}$ to 2.0x10$^{11}$cm$^{-2}$, the anisotropy weakens.  Sample A, at p=1.2x10$^{11}$cm$^{-2}$, shows no significant anisotropy at $\nu$=7/2.}
\end{figure}

In order to further understand the impact of spin-orbit coupling on anisotropic transport in hole systems we have examined a series of 20nm symmetrically doped quantum wells with differing densities.  Figure 3 summarizes the density dependence of the anisotropic transport at $\nu$=7/2 and T $\sim$ 50mK.  In sample A, the transport is isotropic.  With an increase in the density to 2.0x10$^{11}$cm$^{-2}$ in sample B, a nascent anisotropy is visible, but R$_{[01\overline{1}]}$/R$_{[011]}$ only reaches $\sim$ 2 at $\nu$=7/2.  We note that while the amplitude of anisotropy in sample B is small, it displays temperature dependence for 50mK$\leq$T$\leq$150mK similar to that observed in sample C (see Fig. 2).  At a density of 2.3x10$^{11}$cm$^{-2}$ in sample C, the anisotropy becomes more pronounced, with a resistance ratio of about 10.  The data of Figure 3 shows that the ground state at $\nu$=7/2 can undergo an isotropic-to-anisotropic transition as a function of increasing density in 20nm quantum wells.  This density driven transition can also be qualitatively understood.  Our calculations indicated that as the density is decreased there is a quantum phase transition such that the valence LL at $\nu$=7/2 is primarily composed of the N=1 orbital which will stabilize an isotropic state. While our simplistic calculation predicts that this transition occurs at a density of approximately 1.0$\times$10$^{11}$cm$^{-2}$, exchange terms not included in our calculation will likely move this transition to higher density possibly bringing it into better agreement with our experimental observation of isotropic transport in sample A at a density of 1.2$\times$10$^{11}$cm$^{-2}$.  We believe the isotropic behavior observed in sample A is of fundamental origin and is not simply a reflection of an insufficiently clean sample, or insufficiently low measurement temperature.  The mobility of sample A is 2.0$\times$10$^6$cm$^2$/Vs, the highest of all the samples used in this study.  Furthermore, the interaction energy scale determining stripe formation is expected to scale only as p$^{1/2}$.  If the ground sate of sample A were to be anisotropic at lower temperatures, we would have expected to see indications of its onset at T=50mK.  While we cannot rule out the possibility that anisotropy will appear at very low temperatures, we think this scenerio is unlikely.  Finally we noted that the B=0 conductivity ($\sigma$) of sample A is large,  $\sigma$$\gg$e$^2$/h, such that the sample is not expected to exhibit any transport anomalies associated with the metal-to-insulator transition.

We have also examined a single heterojunction (sample D) with density p=2.3x10$^{11}$cm$^{-2}$ to investigate if the anisotropy observed in sample C is stable against a change of the symmetry of the confining potential.
The change in the transport properties is striking.  Even though the density is the same as sample C, the transport in sample D is largely isotropic, especially at $\nu$=7/2.  The observation of a symmetry driven anisotropic-to-isotropic transition at fixed carrier density is a strong indication that spin-orbit coupling alters the ground state at $\nu$=7/2. It is well known that the spin-orbit interaction depends sensitively on the electric field present in quantum well, and not just the 2D density \cite{roland1}.

Our calculations for the single heterojunction indicate the the valence LL at $\nu$=7/2 is comprised of several oscillator functions (N=0 through N=5), but importantly, contains significant contributions from N=0 (13\%) and N=1 (20\%).  While more sophisticated calculations will certainly be necessary to understand in greater detail the nature of anisotropic states in half-filled Landau levels of 2DHSs, the calculated LL structure of the SHJ contrasts sharply with that found in the 20nm symmetric quantum well which does not contain significant contributions from the N=0 and N=1 LLs.
\begin{figure}
\includegraphics[width=\columnwidth]{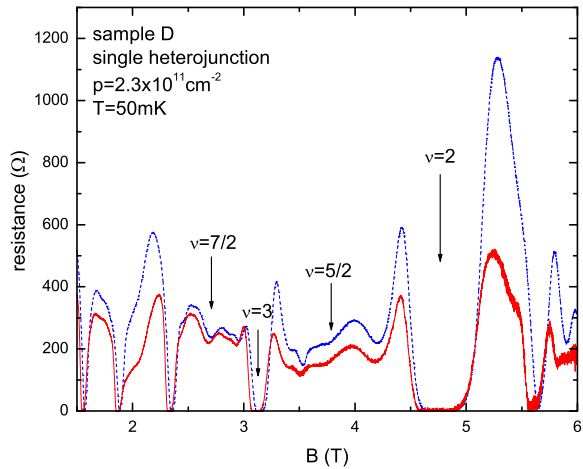}
 \caption{(Color online) Magnetoresistance for sample D at T=50mK.  While the 2D density is the same as sample C, no anisotropy is seen at $\nu$=7/2 in the single heterojunction.}
\end{figure}

In conclusion, we observe
anisotropic transport at filling factors $\nu$=7/2, $\nu$=11/2, and $\nu$=13/2 and isotropic transport at $\nu$=9/2
in  high quality (100) oriented symmetrically doped 2DHSs.  Transport experiments combined with calculations of the Landau level structure indicate that the type of correlated ground state observed at a particular filling factor depends sensitively on the nature of the
single particle states available to the system.

Z.  Jiang is supported by NSF
under DMR-03-52738 and by the DOE under DE-AIO2-04ER46133.

\end{document}